\documentclass[journal]{IEEEtran}

\usepackage{graphicx}
\usepackage{amssymb}
\usepackage{amsfonts}
\usepackage{amsmath}
\usepackage{epsfig}
\usepackage{color}
\usepackage{fancybox}
\usepackage{textcomp}
\usepackage{multirow}
\usepackage{setspace}
\usepackage{psfrag}
\usepackage{color}
\usepackage[ruled,vlined,linesnumbered]{algorithm2e}

\usepackage{hyperref}

\usepackage{mathrsfs}

    \def\Complex{{\rm\rule[.23ex]{.03em}{1.1ex}\kern-.3em{C}}}

    \newcommand{\be}{\begin{equation}} \newcommand{\ee}{\end{equation}}
    \newcommand{\bea}{\begin{eqnarray}} \newcommand{\eea}{\end{eqnarray}}
    \newcommand{\benum}{\begin{enumerate}} \newcommand{\eenum}{\end{enumerate}}

\begin{document}

\title{Model-Driven Deep Learning for Massive Multiuser MIMO Constant Envelope Precoding }

\author{Yunfeng~He, Hengtao~He,~\IEEEmembership{Student Member,~IEEE,} Chao-Kai~Wen,~\IEEEmembership{Member,~IEEE,}\\and~Shi~Jin,~\IEEEmembership{Senior Member,~IEEE}
\thanks{Y.~He, H.~He, and S.~Jin are with the National Mobile Communications Research Laboratory, Southeast University, Nanjing 210096, P. R. China (e-mail: heyunfeng@seu.edu.cn, hehengtao@seu.edu.cn, jinshi@seu.edu.cn).}

\thanks{C.-K. Wen is with the Institute of Communications Engineering, National Sun Yat-sen University, Kaohsiung 804, Taiwan (e-mail:  chaokai.wen@mail.nsysu.edu.tw).}}

\maketitle

\begin{abstract}
Constant envelope (CE) precoding design is of great interest for massive multiuser multi-input multi-output systems because it can significantly reduce hardware cost and power consumption. However, existing CE precoding algorithms are hindered by excessive computational overhead. In this letter, a novel model-driven deep learning (DL)-based network that combines DL with conjugate gradient algorithm is proposed for CE precoding. Specifically, the original iterative algorithm is unfolded and parameterized by trainable variables. With the proposed architecture, the variables can be learned efficiently from training data through unsupervised learning approach. Thus, the proposed network learns to obtain the search step size and adjust the search direction. Simulation results demonstrate the superiority of the proposed network in terms of multiuser interference suppression capability and computational overhead.
\end{abstract}


\section{Introduction}

The massive multiuser multi-input multi-output (MIMO) system has attracted considerable attention because of its superiority in terms of spectral efficiency and reliability~\cite{Lu-An overview}. The base station (BS) utilizes numerous antennas to serve multiple user terminals (UTs) in the same time frequency resource. Linear precoders are usually used to mitigate multiuser interference (MUI) effectively~\cite{Larsson-Massive MIMO}. However, for existing linear precoding algorithms, the actual implementation causes several problems when the number of antennas at the BS is large. A crucial challenge includes the dramatic increase in hardware cost and power consumption. Specifically, each transmit antenna needs to use an expensive linear power amplifier (PA) because the amplitude of the elements in the transmitted signal obtained by existing precoding algorithms, e.g., zero-forcing precoding, is unconstrained.

The type of transmitted signal that facilitates the use of most power-efficient/nonlinear PAs is a constant envelope (CE) signal, i.e., the amplitude of each symbol in the precoding vectors is limited to a constant, and the information is carried on the phase for transmission. Mathematically, the CE precoding design can be formulated as a nonlinear least squares (NLS) problem, which is non-convex and has multiple suboptimal solutions. In~\cite{Mohammend-Per-antenna}, Mohammed and Larsson proposed a sequential gradient descent (GD) algorithm. Unfortunately, this method is greatly affected by the initial value of the iteration algorithm and easily falls into a local minimum, which may reduce the MUI suppression ability dramatically. Reference~\cite{Chen-Improved} proposed a cross-entropy optimization (CEO) method to solve the NLS problem. Although CEO can mitigate MUI effectively, its computational complexity is large, thereby hindering its practical use. In~\cite{Chen-PAPR Precoding}, a Riemannian manifold optimization (RMO)-based conjugate gradient (CG) algorithm that achieves a tradeoff between MUI performance and computational complexity was developed. However, the RMO method still relies on a large number of iterations, which is still a considerable challenge for high-speed communication.

Recently, deep learning (DL) has made remarkable achievements in physical layer communications~\cite{Wang-Deep learning} and has been~introduced into precoding~\cite{Elbir-Hybrid,Xia-A DL framework}.
However, most existing DL-based precoders are designed in a data-driven approach, i.e., considering the precoder as a black-box network, thereby suffering excessively high training cost and computational overhead.
Deep unfolding~\cite{Ito-Trainable, Balatsoukas-Stimming-Deep Unfolding,Monga-Algorithm unrolling} is another DL technique, which expands the iterative algorithms and introduces some trainable parameters to improve the convergence speed, and has been applied to physical layer communications~\cite{He-Model-driven,Balatsoukas-Stimming-NNO}.
In this letter, a model-driven neural network named CEPNet, which combines DL with the RMO-based CG algorithm, is proposed for the CE precoding.
Compared with the RMO-based CG algorithm, the introduced trainable variables can be optimized efficiently through unsupervised learning. Thus, the MUI performance and computational cost of the proposed network have improved significantly.
In addition, simulation results demonstrate that the CEPNet shows strong robustness to channel estimation error and channel model mismatch.

\emph{Notations}---Throughout this letter, we use ${\mathbb{R}}$ and ${\mathbb{C}}$ to denote the set of real and complex numbers, respectively. The superscripts ${{( \cdot )}^{\rm T}}$, ${{( \cdot )}^{\rm H}}$, and ${{( \cdot )}^{*}}$ represent transpose, Hermitian transpose, and conjugate transpose, respectively. ${\circ}$ denotes the Hadamard product between two matrices with identical size. ${\mathfrak{Re}\{\cdot\}}$ returns the real part of its input argument. ${{\|\cdot\|}_{2}}$ and ${|\cdot|}$ represent the Euclidean norm and absolute value, respectively. Finally, for any vector ${\mathbf{z}}$ and any positive integer ${k}$, ${{\left( \mathbf{z} \right)}_{k}}$ returns the $k$th element in vector~${\mathbf{z}}$.

\section{System Model and Problem Formulation}
We consider a downlink MIMO system, in which a BS with $N_{\rm t}$ transmit antennas serves ${N_{\rm u}}$ (${{N}_{\rm u}}<{{N}_{\rm t}}$) single-antenna UTs.
The collectively received signal, denoted by ${\mathbf{y}={{[ {{y}_{1}},~\ldots ,~{{y}_{{{N}_{\rm u}}}}]}^{\rm T}}\in {{\mathbb{C}}^{{{N}_{\rm u}}}}}$, is provided as follows:
\begin{equation}
    \mathbf{y}=\mathbf{Hx}+\mathbf{n},
\end{equation}
where ${\mathbf{H}=[{{h}_{mn}}]\in {{\mathbb{C}}^{{{N}_{\rm u}}\times {{N}_{\rm t}}}}}$ , ${\mathbf{x}={{[{{x}_{1}},~\ldots,~{{x}_{{{N}_{\rm t}}}}]}^{\rm T}}\in {{\mathbb{C}}^{{{N}_{\rm t}}}}}$, and ${\mathbf{n}={{[{{n}_{1}},~\ldots,~{{n}_{{{N}_{u}}}}]}^{\rm T}}\in {{\mathbb{C}}^{{{N}_{\rm u}}}}}$ denote the channel vector, transmitted vector, and additive white Gaussian noise, respectively.
The total MUI energy can be expressed~as
\begin{equation}
	\left\| \mathbf{Hx}-\mathbf{s} \right\|_{2}^{2}\triangleq f\left( \mathbf{x} \right)=\sum\limits_{m=1}^{{{N}_{\rm u}}}{{{\left| \sum\limits_{n=1}^{{{N}_{\rm t}}}{{{h}_{mn}}{{x}_{n}}}-{{s}_{m}} \right|}^{2}}},
\end{equation}
where ${\mathbf{s}={{[{{s}_{1}},~\ldots ,~{{s}_{{{N}_{\rm u}}}}]}^{\rm T}}\in {{\mathbb{C}}^{{{N}_{\rm u}}}}}$ denotes the information symbol vector.

CE precoding, which imposes a constant amplitude constraint on the transmitted signal at each transmit antenna, has been proposed, thereby enabling the utilization of low-cost and high energy efficient PAs.
Mathematically, the design problem that considers CE precoding can be formulated as the constraint optimization problem:
\begin{equation}
\label{eq: Problem}
\begin{aligned}
&\underset{\mathbf{x}}{\mathop{\rm minimize}}\quad f\left(\mathbf{x} \right) \\ 
&{\rm subject~to}~\left|{{x}_{n}}\right|=\sqrt{{{{P}_{\rm t}}}/{{{N}_{\rm t}}}},~{\rm for}~n=1,~\ldots ,~{{N}_{\rm t}},
\end{aligned}
\end{equation}
where ${{P}_{\rm t}}$ denotes the total transmit power. Although no optimized method solves the non-convex problem (\ref{eq: Problem}), the RMO-based CG algorithm can solve the problem with a good trade-off in terms of MUI performance and complexity.

\section{CEPNet}

\subsection{Algorithm Review}
\vspace{-0.1cm}
RMO has been proposed to solve the optimization problem~(\ref{eq: Problem}) in~\cite{Chen-PAPR Precoding}, which transforms the constrained domain into a Riemannian manifold and solves the optimization problem directly on this specific manifold.
We briefly review the RMO-based CG algorithm and recommend~\cite{Absil-Optimization} for technical details about the RMO method.

Considering the constant amplitude constraint for each element of ${\mathbf{x}}$ in problem (\ref{eq: Problem}) and assuming ${P_{\rm t}=1}$, the constraint domain of CE precoding problem can be transformed into a Riemannian manifold given by
\begin{equation} 
	\label{eq: Riemannian manifold}
    \mathcal{M} =\left\{ \mathbf{x}\in {{\mathbb{C}}^{{{N}_{\rm t}}}}:\left| {{x}_{1}} \right|=\left| {{x}_{2}} \right|=\cdots =\left| {{x}_{{N}_{\rm t}}} \right|=\frac{1}{\sqrt{{N}_{\rm t}}} \right\}.
\end{equation}
Given a point ${{\mathbf{x}_{k}} \in \mathcal{M}}$ of the $k$th iteration, the tangent space at point ${\mathbf{x}_{k}}$ is defined as ${\mathcal{T}_{{{\mathbf{x}}_{k}}}}\mathcal{M}=\{\mathbf{z}\in {{\mathbb{C}}^{{{N}_{\rm t}}}}:\mathfrak{Re}\{ \mathbf{z}\circ \mathbf{x}_{k}^{*} \}={{\mathbf{0}}_{{{N}_{\rm t}}}}\}$.
The search step size and direction of the $k$th iteration are assumed as ${{\alpha}_{k}}$ and ${\mathbf{d}_{k}\in \mathcal{T}_{{\mathbf{x}}_{k}}\mathcal{M}}$, respectively.
The point ${\mathbf{x}_{k+1}}$ is obtained by projecting the point ${\mathbf{x}_{k}+{{\alpha}_{k}\mathbf{d}_{k}}}$ back to the manifold as follows:
\begin{equation}
\label{eq: new_x}
	\begin{aligned}
	{{\mathbf{x}}_{k+1}}\hspace{-0.05cm}
		& = {\rm Retr}_{{\mathbf{x}_{k}}}\left({{\alpha}_{k}\mathbf{d}_{k}} \right) \\ 
		& \triangleq \hspace{-0.05cm}\frac{1}{\sqrt{{{N}_{\rm t}}}}\hspace{-0.1cm}\times \hspace{-0.1cm}{{\left[ \frac{{{\left( {{\mathbf{x}}_{k}}+{{\alpha }_{k}}{{\mathbf{d}}_{k}} \right)}_{1}}}{\left| {{\left( {{\mathbf{x}}_{k}}+{{\alpha }_{k}}{{\mathbf{d}}_{k}} \right)}_{1}} \right|},~\ldots ,\frac{{{\left( {{\mathbf{x}}_{k}}+{{\alpha }_{k}}{{\mathbf{d}}_{k}} \right)}_{{{N}_{\rm t}}}}}{\left| {{\left( {{\mathbf{x}}_{k}}+{{\alpha }_{k}}{{\mathbf{d}}_{k}} \right)}_{{{N}_{\rm t}}}} \right|} \right]}^{\rm T}}.
	\end{aligned}
\end{equation}
Next, we introduce how to determine the search direction and step size. Specifically, the CG algorithm is used to determine the search direction.
The gradient direction in Euclidean space is denoted by ${{\nabla f( {{\mathbf{x}}_{k}})=-2{{\mathbf{H}}^{\rm H}}(\mathbf{s}-\mathbf{H}{{\mathbf{x}}_{k}})}}$, which should be projected onto the tangent space at point ${\mathbf{x}_{k}}$~as
\begin{equation}
    \label{eq: projection gradient}
	\begin{aligned}
	{\rm Proj}_{{{\mathbf{x}}_{k}}}\nabla f\left( {{\mathbf{x}}_{k}} \right)
	& \triangleq {\rm grad}f\left( {{\mathbf{x}}_{k}} \right) \\ 
	& =\nabla f\left( {{\mathbf{x}}_{k}} \right)-{{N}_{t}}\times \mathfrak{Re}\left\{ \nabla f\left( {{\mathbf{x}}_{k}} \right)\circ \mathbf{x}_{k}^{*} \right\}\circ {{\mathbf{x}}_{k}}. \\ 
	\end{aligned}	
\end{equation}
Similarly, the search direction ${{\mathbf{d}_{k-1}}\in \mathcal{T}_{{\mathbf{x}}_{k-1}}\mathcal{M}}$ also needs to be projected onto the tangent plane at point ${\mathbf{x}_{k}}$ as
\begin{equation}
	\label{eq: projection search direction}
  	\begin{aligned}
    {\rm Proj}_{{{\mathbf{x}}_{k}}}{{\mathbf{d}}_{k-1}}
    ={{\mathbf{d}}_{k-1}}-{{N}_{t}}\times \mathfrak{Re}\left\{{{\mathbf{d}}_{k-1}} \circ \mathbf{x}_{k}^{*} \right\}\circ {{\mathbf{x}}_{k}}. \\ 
    \end{aligned}
\end{equation}
Then, the search direction ${{\mathbf{d}_{k}}}$ is given by 
\begin{equation}
	\label{eq: new search direction}
    {{\mathbf{d}}_{k}}=-{\rm grad}f\left( {{\mathbf{x}}_{k}} \right)+{{\beta }_{k}}{\rm Proj}_{{{\mathbf{x}}_{k}}}{{\mathbf{d}}_{k-1}},
\end{equation}
where ${{\beta}_{k}}$ is the weight calculated by Polak-Ribière formula as
\begin{equation}
	\label{eq: beta}
	{{\beta }_{k}}=\frac{{\rm grad}f{{\left( {{\mathbf{x}}_{k}} \right)}^{\rm H}}\left( {\rm grad}f\left( {{\mathbf{x}}_{k}} \right)-{\rm Proj}_{{{\mathbf{x}}_{k}}}{\rm grad}f\left( {{\mathbf{x}}_{k-1}} \right) \right)}{{\rm grad}f{{\left( {{\mathbf{x}}_{k-1}} \right)}^{\rm H}}{\rm grad}f\left( {{\mathbf{x}}_{k-1}} \right)}.
\end{equation}
In addition, the search step size ${{\alpha}_{k}}$ restricted to the direction ${{\mathbf{d}_{k}}}$ can be determined according to Armijo backtracking line search rule, that is,
\begin{equation}
    \label{eq: Armijo rule}
	f\left( {{\textbf{x}_{k}} + {{\alpha} _{k}}{\mathbf{d}_{k}}} \right) - f\left( {{\mathbf{x}_{k}}} \right) \le {c_1}{{\alpha} _{k}}{\rm grad}{f^{\rm H}}\left( {{\mathbf{x}_{k}}} \right){\mathbf{d}_{k}},
\end{equation}
where ${{{c}_{1}}\in (0,~1)}$. A large step ${{\alpha}_{k}}$ is usually initialized and attenuated by a factor of ${\tau}$ until (\ref{eq: Armijo rule}) is satisfied. The RMO-based CE precoding method is summarized in Algorithm~\ref{CG-RMO}.
We elaborate the RMO-based CEPNet design in the following subsection.
\begin{algorithm}[t]
	\small
	\label{CG-RMO}
	\caption{RMO-based CG Alg. for CE Precoding }
	\KwIn{${{\mathbf{x}}_{0}\in\mathcal{M}}$}
	\KwOut{Return the final transmitted vector ${{\mathbf{x}}}$}
	{\bf Begin} \\ 
	${{\mathbf{d}}_{0}}=-{\rm grad}f( {{\mathbf{x}}_{0}})$ and ${k=0}$;\\
	{\bf Repeat}\\
	Determine search step size ${{\alpha}_{k}}$ according to (\ref{eq: Armijo rule});\\
	Update the new iterate ${{\mathbf{x}}_{k+1}}$ using the retraction in (\ref{eq: new_x});\\
	Compute Riemannian gradient ${\rm grad}f( {{\mathbf{x}}_{k+1}})$ using (\ref{eq: projection gradient});\\
	Calculate the ${\rm Proj}_{{{\mathbf{x}}_{k+1}}}( {{\mathbf{d}}_{k}})$ using (\ref{eq: projection search direction});\\
	Obtain Polak-Ribière parameter ${{\beta}_{k+1}}$ according to (\ref{eq: beta});\\
	Determine conjugate direction ${{\mathbf{d}}_{k+1}}$ according to (\ref{eq: new search direction});\\
	${k \leftarrow k + 1}$;\\
	{\bf Until} {\emph{Predefined number of iterations is met}};
\end{algorithm}

\subsection{CEPNet Design}
Two primary factors can increase the computational overhead of Algorithm \ref{CG-RMO} significantly. First, the step that determines the search step size ${{\alpha}_{k}}$ takes up almost half the time of one iteration, indicating an excessive and unbearable latency overhead when the number of iterations is large. Second, the projection and retraction operations in Algorithm~\ref{CG-RMO} affect the convergence speed of the CG algorithm, thereby increasing computational complexity.
Therefore, reducing the backtracking line search overhead when determining the step size as much as possible and adjusting the search direction appropriately to accelerate the convergence speed of the CG algorithm are crucial. To this end, we introduce trainable variables to the algorithm and employ DL tools.

Two improvements are applied to the CG algorithm for search step size and search direction:
\begin{enumerate}
	\item Given the ${k}$th iteration, a trainable scalar ${{w}_{k}^{\alpha }\in {{\mathbb{R}}}}$ is defined as the search step size of the iteration ${k}$. All trainable scalars for the search step size constitute the set ${{{\Theta }^{\alpha }}=\{{w}_{k-1}^{\alpha}:k=1,~2,~\ldots ,~K\}}$, where ${K}$ denotes the number of units that represent ${K}$ iterations. Each element of the set ${{\Theta }^{\alpha }}$ is randomly and uniformly initialized between ${{3}\times{10}^{-3}}$ and ${{2}\times{10}^{-2}}$ and trained using the stochastic GD (SGD) algorithm\footnote{The initialization interval is obtained by statistical analysis of the search step size determined by the traditional Armijo backtracking line search rule.}. All trainable variables are fixed after training so that they can be used directly during testing without researching on the basis of the Arimijo backtracking line search rule.
	\item We focus on the weight ${{\beta}_{k}}$ calculated by Polak-Ribière formula in~(\ref{eq: beta}) to adjust the search direction of the iteration ${k+1}$. In particular, we calculate ${{\beta}_{k}}$ according to (\ref{eq: beta}) to obtain a reasonable initial weight value. In addition, a trainable scalar ${{w}_{k}^{\beta }\in {{\mathbb{R}}}}$ is defined and multiplied by ${{\beta}_{k}}$ to determine a new weight factor. All trainable scalars for adjusting the search direction constitute the set ${{{\Theta }^{\beta }}=\{ {w}_{k}^{\beta}:k=1,~2,~\ldots ,~K\}}$. Each element of the set ${{\Theta }^{\beta }}$ is initialized to 1 and trained with SGD. Similarly, the weights are fixed after training.
\end{enumerate}

Finally, we can rewrite~(\ref{eq: new_x}) as
\begin{equation}
\label{eq: weighted alpha}
	{{{\mathbf{x}}_{k+1}}={\rm Retr}_{{\mathbf{x}}_{k}}( {w}_{k}^{\alpha }\times {\mathbf{d}_{k}^{\star}})},
\end{equation}
where ${\mathbf{d}_{k}^{\star}}$ is calculated as
\begin{equation}
\label{eq: weighted beta}
	{\mathbf{d}_{k}^{\star}}=-{\rm grad}f\left( {{\mathbf{x}}_{k}} \right)+{w}_{k}^{\beta }\times {{\beta}_{k}} {\rm Proj}_{{{\mathbf{x}}_{k}}}{{\mathbf{d}}_{k-1}}.
\end{equation}

\begin{figure}
	\centering
	\resizebox{3.5in}{!}{%
	\includegraphics{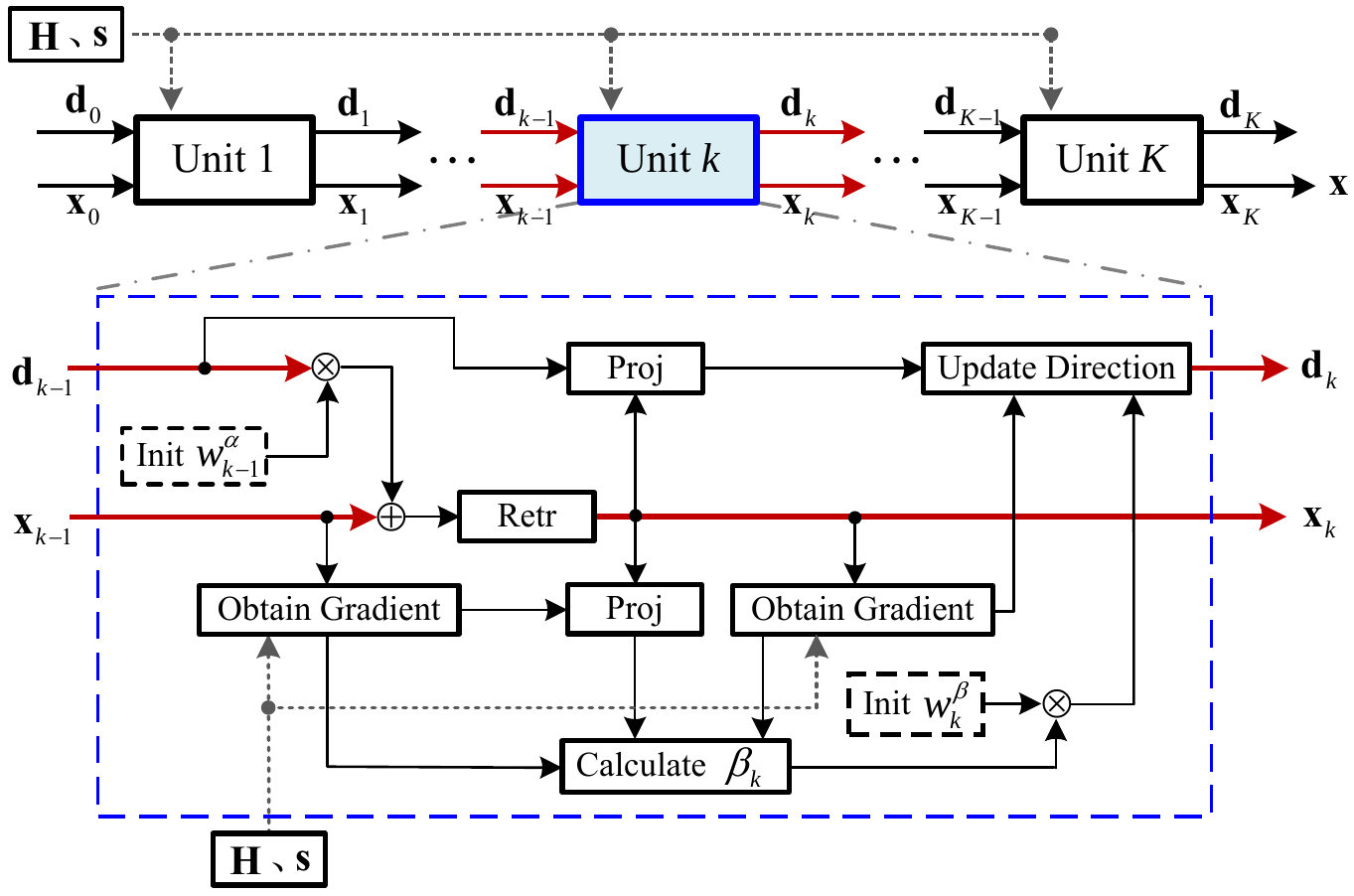} }%
	\caption{Architecture of CEPNet. The CEPNet contains ${K}$ units that represent ${K}$ iterations, and the dashed boxes in each unit denote trainable variables. Apart from replacing~(\ref{eq: new_x}) and~(\ref{eq: new search direction}) by~(\ref{eq: weighted alpha}) and~(\ref{eq: weighted beta}), respectively, the unit undates ${\mathbf{x}_k}$ and ${\mathbf{d}_k}$ according to Algorithm~\ref{CG-RMO}.}
	\label{fig: CEPNet architecture}
\end{figure}

As such, we obtain a CE precoding network named CEPNet, which combines the traditional RMO-based CG algorithm with DL. The overview of the proposed DL architecture is shown in Fig.~\ref{fig: CEPNet architecture}, in which the network inputs are $\mathbf{H}$ and ${\mathbf{s}}$, and the output is ${\mathbf{x}}$. Each unit can be regarded as an iteration of the traditional RMO method. Each unit contains two trainable variables, and the number of total trainable variables in the proposed CEPNet is ${2K}$. Alternatively, each unit contains two active neural layers, and the total number of neural layers in the CEPNet shown in Fig.~\ref{fig: CEPNet architecture} is ${2K}$. On this basis, we do not recommend that the CEPNet contain too many units because excessive neural layers cause the network to be too ``deep,'' which may lead to tricky gradient vanishing or exploding. The proposed network is easy to train because only few trainable variables need to be optimized.

To train CEPNet, supervised learning design is inflexible and inadequate because the optimal label is unknown.
Obtaining labels through existing algorithms is one approach. However, it only makes the network learn the existing algorithms.
Therefore, in our design, an unsupervised learning algorithm is used to train CEPNet effectively. The set of trainable variables is denoted as ${\Theta = \{{\Theta }^{\alpha }, {\Theta }^{\beta } \}}$.
The inputs of CEPNet are $\mathbf{H}$ and ${\mathbf{s}}$, and the transmitted vector is denoted by $\mathbf{x}=g(\mathbf{s};\mathbf{H};\Theta)$.
To improve the robustness of the CEPNet to signal-to-noise ratio (SNR), we use MUI as the loss function directly rather than the average achievable rate or the bit-error rate (BER) that is related to the SNR. Specifically, the loss function is calculated as follows:
\begin{equation}
	\label{eq: loss}
    L(\Theta )=10\times {{\log }_{10}}\left( \frac{1}{{{N}_{\rm s}}{{N}_{{\rm u}}}}\sum\limits_{i=1}^{{N}_{\rm s}}{\left\| {{\mathbf{H}}_{i}}g\left( {{\mathbf{s}}_{i}};{{\mathbf{H}}_{i}};\Theta  \right)-{{\mathbf{s}}_{i}} \right\|_{2}^{2}} \right),
\end{equation}
where ${{N}_{\rm s}}$ denotes the total number of samples in the training set. ${\mathbf{s}_{i}}$ and ${\mathbf{H}_{i}}$ represent the information symbol and channel vectors associated with the \emph{i}th symbol, respectively.

\section{Experiments}
\vspace{0.15cm}
\subsection{Implementation Details}

The CEPNet is constructed on top of the TensorFlow framework, and an NVIDIA GeForce GTX 1080 Ti GPU is used for accelerated training. The training, validation, and testing sets contain $40,000$, $20,000$, and $60,000$ samples, respectively.
We perform simulation experiments with the multipath channel, and the channel vector of the $\mu$th UT is determined by
\begin{equation}
{{\mathbf{h}}^{\mu}}=\frac{1}{\sqrt{L^{\mu}}}\sum\limits_{l=0}^{{{L}^{\mu}}-1}{g_{l}^{\mu} {{\mathbf{a}}^{\mu}}( \theta _{l}^{\mu})},
\end{equation}
where ${L^{\mu}}$ is the number of propagation paths of user $\mu$, and $g_{l}^{\mu}$ denotes the complex gain of the $l$th propagation path in the $\mu$th UT’s channel, and
\begin{equation}
{{{\mathbf{a}}^{\mu}}=\left[ 1,~e^{j2\pi \frac{d}{\lambda }\sin\theta _{l}^{\mu}}, ~\dots,~e^{j2\pi \frac{d}{\lambda }\left({N}_{\rm t}-1\right)\sin\theta _{l}^{\mu}} \right]}^{\rm T}
\end{equation}
denotes the steering vector of the $\mu$th user, where $d$, $\lambda$, and $\theta _{l}^{\mu}$ denote the distance between two horizontally or vertically adjacent antenna elements, the carrier wavelength, and the angle of departure of the $l$th propagation path in the $\mu$th UT's channel, respectively. The channel vector $\mathbf{H}={[{{\mathbf{h}}^{1}},~{{\mathbf{h}}^{2}},~\dots,{{\mathbf{h}}^{N_{\rm u}}}]}^{\rm T}$. We set $L^{\mu}=8$. $g_{l}^{\mu}$ is drawn from ${{\mathcal{N}_\mathbb{C}}(0,~1)}$, and $\theta _{l}^{\mu}$ is uniformly generated between $0$ and $\pi$.
The data sets are formed in ${(\mathbf{s}_i, \mathbf{H}_i)}$ pair. The channel vector ${{\mathbf{H}}_{i}}$ is set as block fading, and one transmitted vector ${{\mathbf{s}}_{i}}$ corresponds to one channel vector ${{\mathbf{H}}_{i}}$, where each ${{\mathbf{H}}_{i}}$ is generated independently and each element of the transmitted vector ${{\mathbf{s}}_{i}}$ is drawn from the 16-QAM constellation.
We set ${K=20}$ for the trade-off between MUI performance and computational complexity. The set of trainable variables is updated by the ADAM optimizer~\cite{Kingma-Adam}. The training epochs, learning rate, and batch size are set as $500$, $0.0002$, and $4,000$, respectively.

\subsection{Performance Analysis}
\label{Performance Analysis}
\subsubsection{Average achievable rate}
\begin{figure}
	\centering
	\resizebox{3.6in}{!}{
	\includegraphics*{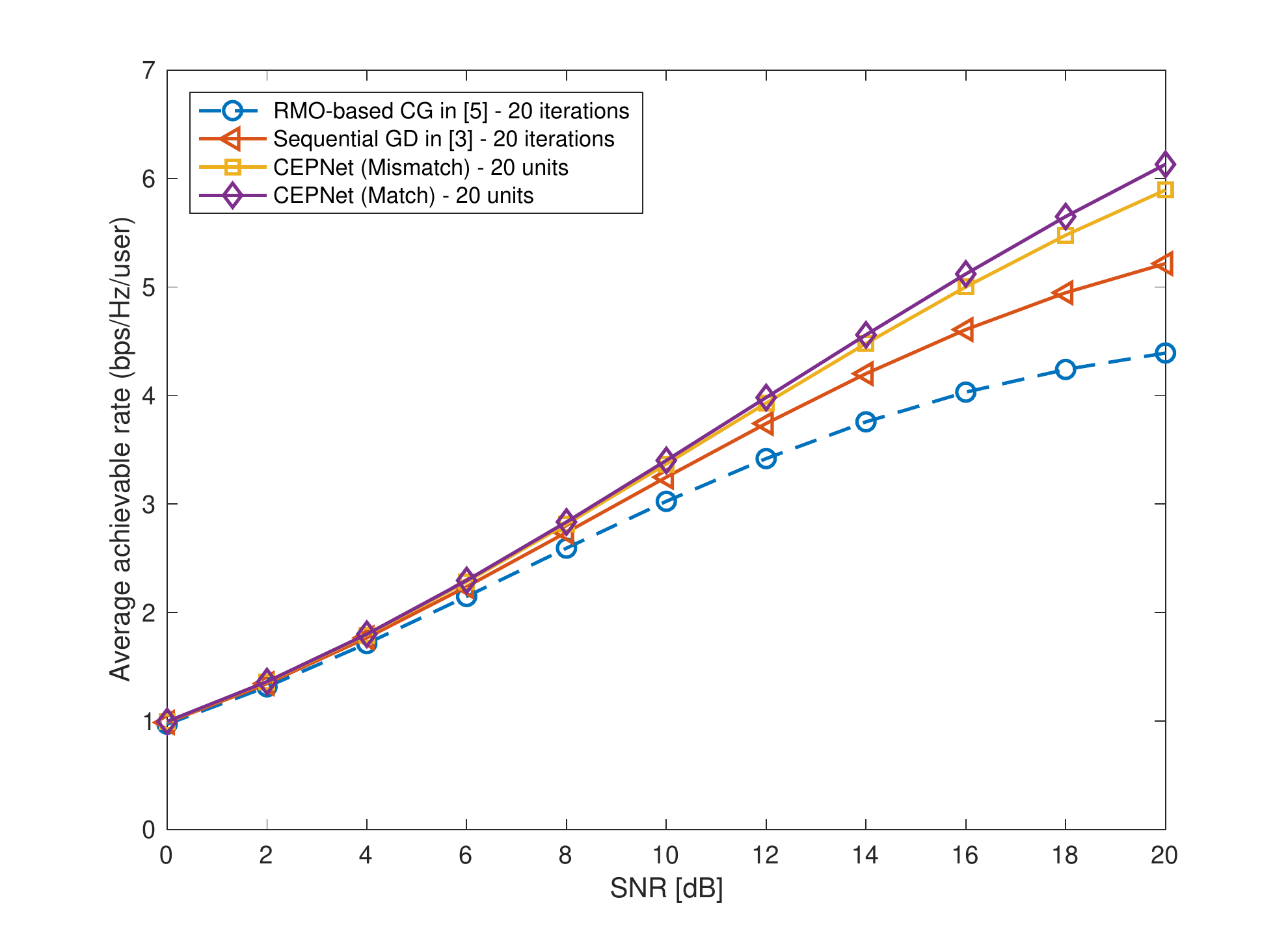}}
	\caption{\label{fig: average achievable rate sparse-path}Average achievable rate against the SNR in the multipath channel, where $N_{\rm t}=64$ and $N_{\rm u}=16$. The numbers of iterations for the RMO-based CG, the sequential GD, and the CEPNet are all 20.}
\end{figure}

In Fig.~\ref{fig: average achievable rate sparse-path}, we compare the performance of the proposed CEPNet with the existing CE precoding algorithms on the average achievable rate against the SNR in a multipath channel, where the achievable rate at each UT can be calculated using~\cite[Eq.~(43)]{Mollen-Waveforms}. The BS is equipped with ${{N}_{\rm t}=64}$ transmit antennas and serves ${{N}_{\rm u}=16}$ UTs.
Fig.~\ref{fig: average achievable rate sparse-path} indicates that the CEPNet trained with the matched multipath channel outperforms the RMO-based CG algorithm with the same number of iterations significantly, demonstrating that the proposed CEPNet can learn to reduce the total MUI through an unsupervised learning approach, i.e., the CEPNet learns to adjust the search step size and direction of each iteration appropriately through training. In addition, the CEPNet also outperforms the sequential GD algorithm with the same number of iterations at high SNRs. We infer that the CEPNet learns to deal with the channel singular issue in the multipath channel through training, while the sequential GD algorithm fails. In general, the average achievable rate performance of the CEPNet is better than the existing CE precoding algorithms in the multipath channel.

\subsubsection{Bit-error rate}
\begin{figure}
	\centering
	\resizebox{3.5in}{!}{
		\includegraphics*{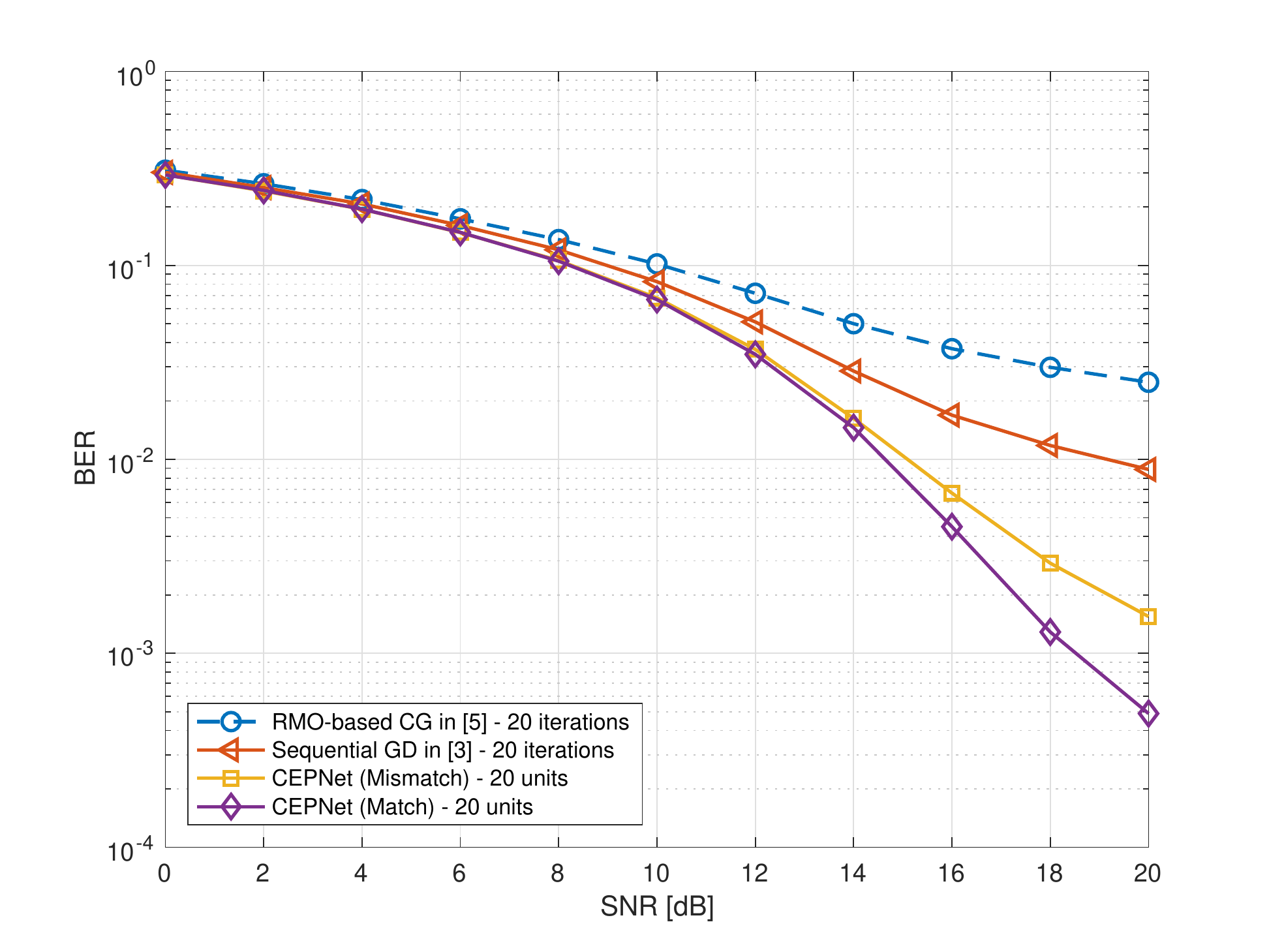}}
	\caption{\label{fig: ber}BER against the SNR in the multipath channel, where $N_{\rm t}=64$ and $N_{\rm u}=16$. The numbers of iterations for the RMO-based CG, the sequential GD, and the CEPNet are all 20.}
\end{figure}

Fig.~\ref{fig: ber} compares the BER performance of the CEPNet with existing CE precoding algorithms in the multipath channel, where the BS is equipped with ${{N}_{\rm t}=64}$ transmit antennas and serves ${{N}_{\rm u}=16}$ UTs.
In Fig.~\ref{fig: ber}, the CEPNet trained with the matched channel model obtains the best BER performance among all CE precoders. Specifically, the CEPNet outperforms the RMO-based CG algorithm with the same number of iterations by approximately $5.8$~dB when we target SNR for $\text{BER}=0.03$. Similarly, the CEPNet outperforms the sequential GD algorithm with the same number of iterations by approximately $4.5$~dB when we target SNR for $\text{BER}=0.01$.

\subsubsection{Computational complexity}
We compare the computational overhead of the RMO-based CG algorithm, the sequential GD algorithm, and the CEPNet\footnote{We first construct the CEPNet on top of the TensorFlow framework and an NVIDIA GeForce GTX 1080 Ti GPU is used for accelerated training. Once the CEPNet converges, the parameters are stored into .csv files. The parameters of the CEPNet are obtained from the pre-stored files directly when we implement the CEPNet with MATLAB framework.} with same number of iterations in Table~\ref{table: computational overhead}, where $N_{\rm t}=64$ and $N_{\rm u}=16$. The aforementioned CE precoders are all implemented with MATLAB framework for fairness. Time comparison is performed on a computer with OSX 10.12, i5-6360U 2.9~GHz dual-core CPU, and 8~GB RAM.
\begin{table}
\centering
\caption{The comparison of the computational overhead}
\label{table: computational overhead}
\begin{footnotesize}

\begin{tabular}{c|c}
	\hline
	CE precoding algorithms & time (in seconds) \\
	\hline\hline
	RMO-based CG -- 20 iterations & 0.00079 \\
	\hline
	Sequential GD in~\cite{Mohammend-Per-antenna} -- 20 iterations & 0.0321 \\
	\hline
	Our proposed CEPNet -- 20 units & 0.00036 \\
	\hline
\end{tabular}
\end{footnotesize}
\end{table}
The results indicate that CE precoding through the CEPNet can be executed with a lower overhead than that through the RMO-based CG algorithm because the former does not require any backtracking on the search step size. Specifically, the CEPNet with $20$ units performs approximately $2.19$ and $89.17$ times faster than the RMO-based CG algorithm with $20$ iterations and the sequential GD algorithm with $20$ iterations, respectively.

\subsection{Robustness Analysis}
\vspace{0.05cm}
\subsubsection{Robustness to channel estimation error}
In Sec.~\ref{Performance Analysis}, we assume that the BS can obtain perfect channel state information (CSI) for precoding. However, considering that obtaining perfect CSI is impractical, we first investigate the robustness of the CEPNet to channel estimation error in this section. The channel estimation vector $\hat{\mathbf{H}}_{i}$ used for precoding is assumed to be given by
\begin{equation}
	\hat{\mathbf{H}}_{i}=\sqrt{1-\varepsilon }\mathbf{H}_{i}+\sqrt{\varepsilon }\mathbf{E}_{i},
\end{equation}
where $\epsilon \in[0,~1]$ and $\mathbf{E}_{i}$ is drawn from ${{\mathcal{N}_\mathbb{C}}(0,~1)}$. The value of $\epsilon$ measures the magnitude of the channel estimation error. The CEPNet is trained with perfect CSI. We evaluate the RMO-based CG algorithm, the sequential GD algorithm, and the CEPNet with imperfect CSI. The aforementioned CE precoding algorithms are all performed with $20$ iterations and $\text{SNR}=20$~dB.
\begin{figure}
	\centering
	\resizebox{3.5in}{!}{
	\includegraphics*{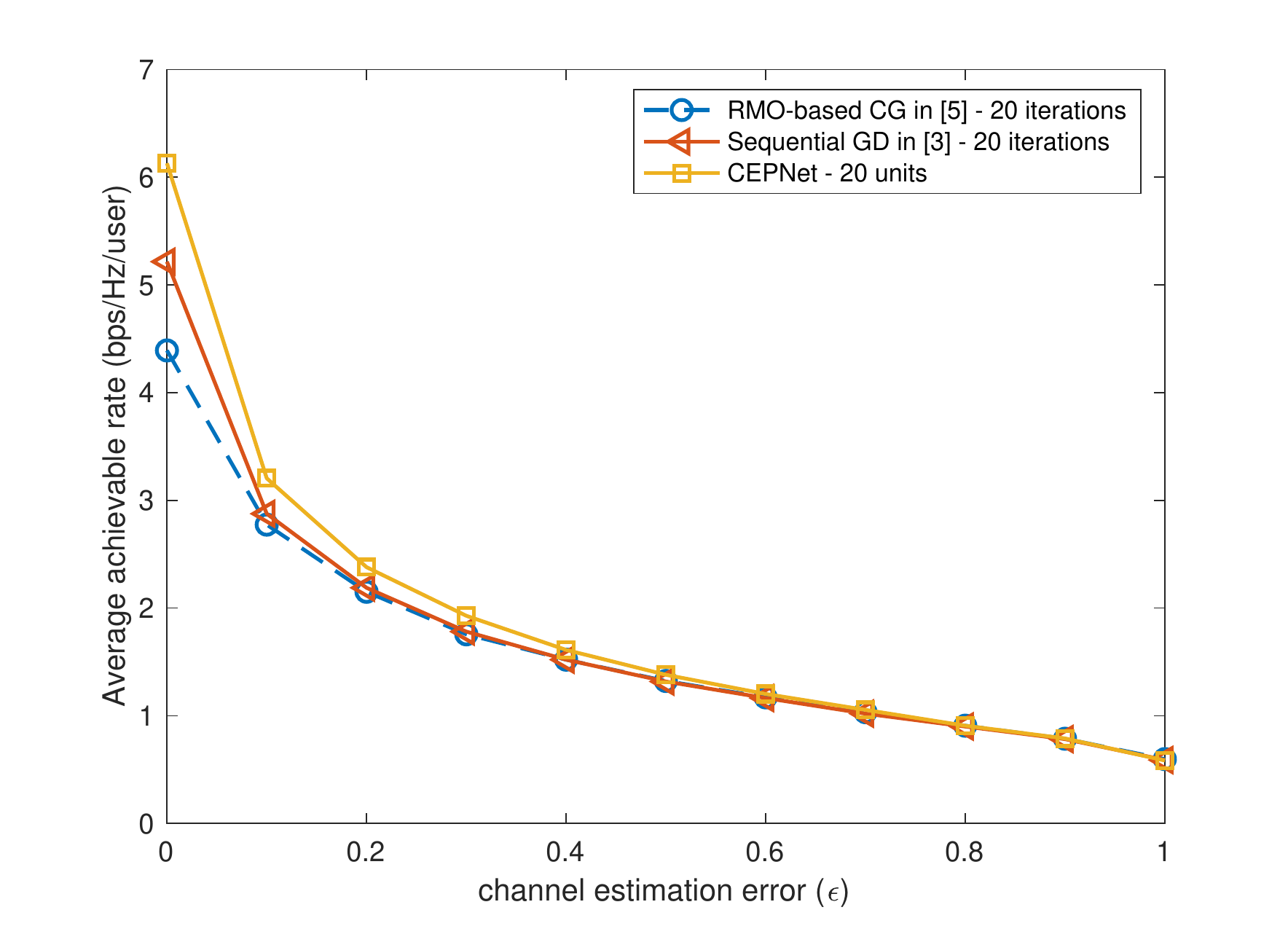}}
	\caption{\label{fig: roubustness}Average achievable rate performance of the RMO-based CG algorithm, the sequential GD algorithm, and the CEPNet with the channel estimation error $\epsilon$ in a multipath channel with $\text{SNR}=20$~dB, where $N_{\rm t}=64$ and $N_{u}=16$. The numbers of iterations for different CE precoding algorithms are all 20.}
\end{figure}

Fig.~\ref{fig: roubustness} illustrates the robustness of different CE precoding algorithms to the channel estimation error. The figure shows that the proposed CEPNet outperforms the RMO-based CG algorithm and the sequential GD algorithm when $\epsilon \in[0,~0.5]$, which indicates that the learned variables ${\Theta}$ are robust to channel estimation error. In addition, the performance of the aforementioned CE precoding algorithms is similar when $\epsilon \in (0.5,~1]$ because the channel estimation error is significant.
\vspace{0.05cm}
\subsubsection{Robustness to channel model mismatch}
We investigate the robustness of the CEPNet to channel model mismatch. Specifically, the CEPNet is trained with a Rayleigh-fading channel and deployed with a multipath channel. Each element of the Rayleigh-fading channel is drawn from ${{\mathcal{N}_\mathbb{C}}(0,~1)}$. Figs.~\ref{fig: average achievable rate sparse-path} and~\ref{fig: ber} indicate that the CEPNet trained with the Rayleigh-fading channel can also achieve significant gains compared with existing CE precoders in the multipath channel, which demonstrates that the learned variables are robust to channel model mismatch.

In general, the CEPNet shows strong robustness to channel estimation error and channel model mismatch. We also do not have to retrain the CEPNet if the channel variation is insignificant.
However, we should retrain the CEPNet to improve performance when the channel variation amplitude is significant. Considering that the CEPNet only contains the $2K$ parameter, we can retrain the CEPNet with a low overhead to adapt the changed channel.

\section{Conclusion}
We proposed a novel model-driven DL network for multiuser MIMO CE precoding. The designed CEPNet inherited the superiority of the conventional RMO-based CG algorithm and DL technology, thereby exhibiting excellent MUI suppression capability.
Simulation results demonstrated that the CEPNet could reduce the precoding overhead significantly compared with the existing CE precoding algorithms.
Furthermore, the CEPNet showed strong robustness to the channel estimation error and the channel model mismatch.

{\renewcommand{\baselinestretch}{1.05}
\begin{footnotesize}
\bibliographystyle{IEEEtran}


\end{footnotesize}}

\end{document}